\documentclass[
aps,
prd,
showpacs,
preprint,
tightenlines,
nofootinbib,
floatfix]
{revtex4}
\usepackage{graphicx,
}
\begin{document}
%
\title{                Evidence for Mikheyev-Smirnov-Wolfenstein effects\\
                           in solar neutrino flavor transitions}
%
\author{        G.L.~Fogli}
\affiliation{   Dipartimento di Fisica
                and Sezione INFN di Bari\\
                Via Amendola 173, 70126 Bari, Italy\\}
\author{        E.~Lisi}
\affiliation{   Dipartimento di Fisica
                and Sezione INFN di Bari\\
                Via Amendola 173, 70126 Bari, Italy\\}
\author{        A.~Marrone}
\affiliation{   Dipartimento di Fisica
                and Sezione INFN di Bari\\
                Via Amendola 173, 70126 Bari, Italy\\}
\author{        A.~Palazzo}
\affiliation{   Dipartimento di Fisica
                and Sezione INFN di Bari\\
                Via Amendola 173, 70126 Bari, Italy\\}

\begin{abstract}
We point out that the recent data from the Sudbury Neutrino
Observatory, together with other relevant measurements from solar
and reactor neutrino experiments, convincingly show that the
flavor transitions of solar neutrinos are affected by
Mikheyev-Smirnov-Wolfenstein (MSW) effects. More precisely, one
can safely reject the null hypothesis of no MSW interaction energy
in matter, despite the fact that the interaction amplitude
(formally treated as a free parameter) is still weakly constrained
by the current phenomenology. Such a constraint can be improved,
however, by future data from the KamLAND experiment. In the
standard MSW case, we also perform an updated analysis of
two-family active oscillations of solar and reactor neutrinos.
\end{abstract}
\medskip
\maketitle

\section{Introduction}

The Sudbury Neutrino Observatory (SNO) experiment has recently
released new data \cite{Hott} with enhanced sensitivity to
neutral-current (NC) interactions of solar neutrinos in deuterium.
Charged current (CC) and elastic scattering (ES) events have also
been statistically separated from NC events in a model-independent
way, i.e., without using priors on the $^8$B neutrino energy
spectrum shape \cite{Hott}. These data corroborate the explanation
of the solar neutrino deficit \cite{Book} in terms of (dominant)
two-family $\nu_e\to\nu_a$ flavor transitions
($\nu_a=\nu_{\mu,\tau}$), which have convincingly emerged from the
combined data of previous solar neutrino experiments (Chlorine
\cite{Cl98}, Gallium \cite{Ab02,Ha99,Ki02}, Super-Kamiokande (SK)
\cite{Fu01,Fu02}, and SNO \cite{AhCC,AhNC,AhDN}) and of
long-baseline reactor oscillation searches at KamLAND \cite{KamL}.
Moreover, the new SNO data appear to forbid relatively high values
of the neutrino mixing angle $\theta_{12}$ (close to maximal
mixing) and of the squared mass difference $\delta m^2$ (close to
the CHOOZ \cite{CHOO} upper bound), which were marginally allowed
prior to \cite{Hott} (see, e.g., \cite{Our1,Pena}). In the current
global fit \cite{Hott}, the mass-mixing parameters appear to be
tightly confined in the so-called large mixing angle (LMA) region,
and especially in a subregion previously denoted as LMA-I
\cite{Our1}.

In the LMA parameter range, flavor transitions between $\nu_e$ and
$\nu_a$ should be significantly affected by the neutrino
interaction energy difference $V=V_e-V_a$ arising in solar (and
possibly Earth) background matter \cite{Matt,Barg},
\begin{equation}\label{V}
  V(x)=\sqrt{2} G_F N_e(x)\ ,
\end{equation}
where $N_e$ is the electron number density at the point $x$. The
associated flavor change, known as Mikheyev-Smirnov-Wolfenstein
(MSW) effect \cite{Matt}, should occur adiabatically \cite{Adia}
in the solar matter, for LMA parameters. In the context of
Hamiltonian ($\cal H$) evolution of $2\nu$ active flavors, the MSW
effect enters through a dynamical term ${\cal H}_\mathrm{dyn}$ in
matter, in addition to the kinetic term ${\cal H}_\mathrm{kin}$ in
vacuum:
\begin{equation}\label{H}
  i\frac{d}{dx}\left(\begin{array}{c}\nu_e\\ \nu_a\end{array}\right)=
  ({\cal H}_\mathrm{dyn}+{\cal H}_\mathrm{kin})
  \left(\begin{array}{c}\nu_e\\ \nu_a\end{array}\right)\ ,
\end{equation}
where
\begin{equation}
  \label{Hdyn}
  {\cal H}_\mathrm{dyn}=\frac{V(x)}{2}
  \left(\begin{array}{cc} 1 & 0\\
  0&-1\end{array}\right)\
\end{equation}
and
\begin{equation}
\label{Hkin}
  {\cal H}_\mathrm{kin} = \frac{\delta m^2}{4 E}
  \left(\begin{array}{cc}-\cos2\theta_{12}&\sin2\theta_{12}\\
  \sin2\theta_{12}&\cos2\theta_{12}\end{array}\right)\ ,
\end{equation}
$E$ being the neutrino energy.

In a previous work \cite{MSW1} we pointed out that, while the
evidence for ${\cal H}_\mathrm{kin}\neq 0$ was overwhelming, the
phenomenological indications in favor of ${\cal
H}_\mathrm{dyn}\neq 0$ (and thus of MSW effects) were not as
compelling. In particular, we introduced in \cite{MSW1} a free
parameter $a_\mathrm{MSW}$ modulating the overall amplitude of the
dynamical term ${\cal H}_\mathrm{dyn}$ through the substitution
\begin{equation}
\label{Ha} V \to a_\mathrm{MSW}\cdot V\ ,
\end{equation}
both in the Sun and in the Earth. We showed that $a_\mathrm{MSW}$
was poorly constrained, despite an intriguing preference for the
standard MSW expectation $a_\mathrm{MSW}\sim 1$ \cite{MSW1}. The
null hypothesis $a_\mathrm{MSW}=0$ was not clearly disproved by
any single experiment, and could be rejected at a relevant
confidence level ($\Delta\chi^2\simeq 13$, formally equivalent to
$\sqrt{\Delta\chi^2}\simeq3.5\sigma$) only in the global fit. We
concluded that the available phenomenological data clearly favored
MSW effects in solar neutrinos, but did not prove unequivocally
their occurrence. We deemed it necessary to wait for new KamLAND
or SNO data, in order to clarify the situation and to probe MSW
effects with higher statistical significance \cite{MSW1}.

In this work, we point out that the recent SNO data \cite{Hott}
contribute significantly to disprove the null hypothesis of no MSW
oscillations. In the global combination of solar and reactor data,
we find that, with respect to the (preferred) standard case
$a_\mathrm{MSW}\sim 1$, the null hypothesis $a_\mathrm{MSW}=0$ can
be safely rejected at the level of $\sim 5.6\sigma$, despite the
fact the allowed range of $a_\mathrm{MSW}$ is still rather large.
In other words, the evidence in favor of MSW effects is now very
strong, although precision tests of the MSW physics cannot be
performed until new, high statistics KamLAND data become available
(as we show later).

In the following sections, we analyze the current solar and
reactor neutrino phenomenology with an increasing degree of
dependence on assumptions about the MSW effect.%
\footnote{In any case, we assume active flavor oscillations only,
and do not consider hypothetical sterile neutrinos. }
In Sec.~II we do not make any hypothesis about MSW effects, and
show that SNO data alone, as well as a model-independent SNO+SK
combination, constrain the energy-averaged $\nu_e$ survival
probability $\langle P_{ee}\rangle$ to be significantly smaller
than $1/2$. This fact, by itself, excludes the vacuum case
$a_\mathrm{MSW}=0$ (which would predict $\langle P_{ee}\rangle
\geq 1/2$ in the LMA region selected by KamLAND), and proves that
dynamical effects {\em must\/} occur in solar neutrino propagation
with unspecified amplitude $a_\mathrm{MSW}> 0$. In Sec.~III we fit
all the available solar and reactor data with ($\delta
m^2,\theta_{12},a_\mathrm{MSW}$) taken as free parameters. We find
that MSW effects with standard amplitude ($a_\mathrm{MSW}=1$) are
favored, while the null hypothesis $(a_\mathrm{MSW}=0)$ can be
safely rejected at the $\sim5.6\sigma$ level. However, we show
that the allowed range of $a_\mathrm{MSW}$ is still very large,
and can be significantly narrowed only by future KamLAND data.
Assuming standard MSW effects $(a_\mathrm{MSW}=1)$, we perform in
Sec.~IV  an updated analysis of  the $2\nu$ kinematical parameters
$(\delta m^2,\sin^2\theta_{12})$. We briefly discuss the impact of
$3\nu$ mixing in Sec.~V, and conclude our work in Sec.~VI.

\section{Model-independent constraints}

It has been shown in \cite{Vill} (see also \cite{Gett}) that the
SK and SNO experiments probe the same energy-averaged $\nu_e$
survival probability $\langle P_{ee}\rangle$ to a good accuracy,
provided that the detector thresholds are appropriately chosen.
For the kinetic energy threshold ($T_\mathrm{SNO}=5.5$ MeV) and
energy resolution characterizing the latest SNO data \cite{Hott},
we find that
the equivalent SK threshold is $E_\mathrm{SK}\simeq 7.8$ MeV in total energy.%
\footnote{In practice, adopting this threshold amounts to discard
the first three bins of the SK energy spectrum in the range $E\in
[5,8]$ MeV. }
For equalized thresholds,  the SK ES flux and the SNO NC and CC
fluxes are linked by the {\em exact\/} relations \cite{Vill}
\begin{eqnarray}
\Phi_\mathrm{ES}^\mathrm{SK} &=& \Phi_B [\langle P_{ee}\rangle + r
(1-\langle P_{ee}\rangle)] \ ,\label{phiessk}\\
\Phi_\mathrm{CC}^\mathrm{SNO} &=& \Phi_B \langle P_{ee}\rangle\ ,\label{phiccsno}\\
\Phi_\mathrm{NC}^\mathrm{SNO} &=& \Phi_B \ ,\label{phincsno}
\end{eqnarray}
where $r=0.154$ is the ratio of (properly averaged)
$\nu_{\mu,\tau}$ and $\nu_e$ CC cross sections, and   $\Phi_B$ is
the true $^8$B flux from the Sun. From the above equations, one
can (over)constrain both $\Phi_B$ and $\langle P_{ee}\rangle$ in a
truly model-independent way, namely, without any prior assumption
about the energy profile of $P_{ee}$ or about $\Phi_B$ predictions
in standard solar models (SSM).

Figure~1 shows the current constraints on $\Phi_B$ and on $\langle
P_{ee}\rangle$ as derived from the  final SK ES data \cite{Fu02}
and from the latest SNO CC and NC fluxes \cite{Hott} (correlations
included \cite{Howt}). The constraints are shown both by
individual bands and by their combination at the $3\sigma$ level
$(\Delta \chi^2=9)$. The projections of the SNO+SK combination
(solid ellipse in Fig.~1) provide the ranges
\begin{equation}
\label{PhiB}  \Phi_{B}  = (5.5\pm 1.2)\times 10^6\mathrm{\
cm}^{-2}\mathrm{s}^{-1}  \ (3\sigma)\ ,
\end{equation}
in good agreement with  SSM predictions \cite{BP00}, and
\begin{equation}
\label{Pee} \langle P_{ee} \rangle = 0.31^{+0.12}_{-0.08} \
(3\sigma)\ .
\end{equation}
The above $3\sigma$ limits on $\langle P_{ee}\rangle$ are in very
good agreement with the ``$3\sigma$ range'' obtained by naively
triplicating the errors of the SNO CC/NC flux ratio, which is a
direct measurement of $\langle P_{ee}\rangle$:
$\Phi_\mathrm{CC}^\mathrm{SNO}/\Phi_\mathrm{NC}^\mathrm{SNO}=0.306\pm0.105
(3\sigma)$ \cite{Hott}. However, as emphasized in \cite{Howt}, the
errors of the CC/NC ratio are not normally distributed, and should
not be used in fits. Conversely, our bounds in Eq.~(\ref{Pee}) are
statistically safe and well-defined, and will be used in the
following discussion.

The above SK+SNO constraints appear to be currently dominated by
the SNO data, as shown by the dotted ellipse in Fig.~1. In
particular, the upper bound on the $\nu_e$ survival probability,
\begin{equation}
\label{Peeupper}
 \langle P_{ee} \rangle < 0.43 \ (3\sigma)\ ,
\end{equation}
can be basically derived from the SNO (CC+NC) data \cite{Hott}
alone. The upper limit in Eq.~(\ref{Peeupper}) is significantly
stronger than the one derived in \cite{Gett} prior to the latest
SNO data \cite{Hott}. In particular, we have now robust,
model-independent evidence that $P_{ee}$ is definitely smaller
than 1/2 at $>3\sigma$ level. This inequality has important
consequences for both the dynamical and the kinematical term in
Eq.~(\ref{H}). First, in the $\delta m^2$ range accessible to
KamLAND and below the CHOOZ bound ($\delta m^2\sim O(10^{-4\pm
1})$ eV$^2$), the absence of the dynamical MSW term ${\cal
H}_\mathrm{dyn}$ (i.e., the case $a_\mathrm{MSW}=0$ would imply
$\langle P_{ee}\rangle \geq 1/2$ (see, e.g, \cite{MSW1}), contrary
to Eq.~(\ref{Peeupper}) . Second, assuming standard MSW dynamics
($a_\mathrm{MSW}=1$), the inequality in Eq.~(\ref{Peeupper})
allows to place upper limits on the kinematical parameters $\delta
m^2$ and $\sin^2\theta_{12}$ (see, e.g., the discussions in
\cite{MSW1,PeMa,Pedr}).

Figure~2 illustrates  the previous statements through isolines of
$\langle P_{ee}\rangle$ in the $(\delta m^2,\sin^2\theta_{12})$
parameter range relevant to KamLAND, for both $a_\mathrm{MSW}=0$
(left panel) and $a_\mathrm{MSW}=1$ (right panel). The superposed
grey region is allowed by the $3\sigma$ model-independent bounds
in Eq.~(\ref{Pee}). No such region exists in the case
$a_\mathrm{MSW}=0$, which is therefore rejected at the $3\sigma$
level (at least). In the standard MSW case (left panel), the
allowed region appears to confine $\delta m^2$ below $\sim 2\times
10^{-4}$ eV$^2$ and $\sin^2\theta_{12}$ below $\sim 0.4$. In
particular, the latest SNO data significantly contribute to reject
maximal mixing $(\sin^2\theta_{12}=1/2)$ and to reduce the
likelihood of the so-called \cite{Our1} LMA-II parameter region,
as emphasized in \cite{Hott}.

Summarizing, the latest SNO CC and NC data \cite{Hott}, either by
themselves or in combination with the SK ES data \cite{Fu02},
provide the strong, model-independent upper bound $\langle
P_{ee}\rangle<0.43$ at $3\sigma$. In the context of $2\nu$ mixing,
and within the mass-mixing region probed by KamLAND, this bound
allows to reject the null hypothesis ($a_\mathrm{MSW}=0$), and
provides upper limits on the mass-mixing parameters in the
standard MSW case ($a_\mathrm{MSW}=1$). In the next Section, we
examine the more general case of variable $a_\mathrm{MSW}$, in
order to test whether current and future data can significantly
constrain, by themselves, the size of matter effects.

\section{Constraints on the MSW dynamical term}

In this section we present the results of a global analysis of
solar and reactor (KamLAND+CHOOZ) data with $(\delta
m^2,\sin^2\theta_{12},a_\mathrm{MSW})$ unconstrained. The latest
SNO data \cite{Hott} are incorporated according to the
recommendations in \cite{Howt}. The reader is referred to
\cite{MSW1} for other details of the analysis.

Figure~3 shows the results of the global $\chi^2$ fit, in terms of
the function $\Delta\chi^2(a_\mathrm{MSW})$ after $(\delta
m^2,\sin^2\theta_{12})$ marginalization. Such marginalization is
appropriate to test the size of ${\cal H}_\mathrm{dyn}$
independently of ${\cal H}_\mathrm{kin}$. It can be seen that the
best fit is intriguingly close to the standard case
$(a_\mathrm{MSW}=1)$, although there are other acceptable local
minima over about three decades in $a_\mathrm{MSW}$. As discussed
in \cite{MSW1} for the case of variable $a_\mathrm{MSW}$, the
$\delta m^2$ range allowed by solar neutrino data sweeps through
the tower of LMA-$n$ solutions allowed by KamLAND, leading to a
series of ``bumps'' in the $\Delta\chi^2$ function (solid line).
Such features are unavoidable, as far as KamLAND allows multiple
solutions in the mass-mixing parameter space. However, the
situation should improve with higher KamLAND statistics. Assuming
that KamLAND will confirm the current best-fit solution in the
$(\delta m^2,\sin^2\theta_{12},a_\mathrm{MSW})$ space, and
simulating the corresponding KamLAND data, we obtain the
prospective dotted and dahed curves in Fig.~1, which refer,
respectively, to a fivefold and tenfold increase of the present
statistics (54 events \cite{KamL}). It appears that, with the help
of a few hundreds KamLAND events, the global fit of solar and
reactor data can pinpoint the predicted size of MSW effects within
a factor of $\sim 2$, allowing future ``precision tests'' of this
effects (e.g., to probe additional nonstandard interactions).

Although the current bounds on $a_\mathrm{MSW}$ appear to be
rather weak, the rejection of the null hypothesis
$a_\mathrm{MSW}=0$ is quite strong, and corresponds to a
significance level of $\Delta\chi^2\simeq 32$, i.e.,
$\sim\!5.6\sigma$. Summarizing the results of this and the
previous section, we can state that current solar and reactor data
reject the hypothesis of no MSW effect at $>5\sigma$ level, with a
$>3\sigma$ contribution from the recent SNO data \cite{Hott}.
Therefore, in our opinion, the phenomenological indications in
favor of MSW effects can now be promoted to the level of evidence.

\section{Constraints on kinematical mass-mixing term}

In this section, assuming standard MSW dynamics, we update our
previous bounds \cite{Our1} on the mass-mixing parameters $(\delta
m^2,\sin^2\theta_{12})$ which govern the kinematical term ${\cal
H}_\mathrm{kin}$. The reader is referred to \cite{Our1,Gett} for
technical details. Here we just add that the statistical
correlations of recent SNO data \cite{Hott} are incorporated
through a straightforward generalization of the pull approach
\cite{Gett}, as explicitly described in \cite{Bala}. We have
checked that our analysis ``SNO data only'' reproduces the results
of \cite{Hott} with very good accuracy. Finally, we have updated
the total rate and winter-summer asymmetry from Gallium
experiments \cite{Lown}. In total, we have 84 solar neutrino
observables, plus 13 KamLAND bins.

Figure~4 shows the results of our fit to all solar neutrino data.%
\footnote{We used to add CHOOZ data \cite{CHOO} in order to
strengthen the upper bound on $\delta m^2$ \cite{Our1,Gett}.
However, current solar neutrino data make this addition no longer
necessary in the context of $2\nu$ mixing with standard MSW
effects.}
The best fit $(\chi^2_{\min}=72.9)$ is reached at $\delta
m^2=5.7\times 10^{-5}$ and $\sin^2\theta_{12}=0.29$.  The upper
and lower bounds on the mass-mixing parameters are in good
agreement with the results in \cite{Hott}, and confirm that the
solar neutrino parameter space is steadily narrowing.

Figure~5 incorporates the analysis of KamLAND data \cite{KamL} as
in \cite{Our1}. The best fit $(\chi^2_{\min}=79.7)$ is reached at
$\delta m^2=7.2\times 10^{-5}$ and $\sin^2\theta_{12}=0.29$ (LMA-I
solution), while the second best fit (LMA-II solution) is only
marginally allowed at the $\Delta \chi^2=9.4$ level ($\sim 99\%$
C.L.\ for $N_\mathrm{DF}=2$). Also in this case, we find good
agreement with the results in \cite{Hott}, modulo the obvious
transformation from our linear abscissa $\sin^2\theta_{12}$ to
their logarithmic abscissa $\tan^2\theta_{12}$.

In conclusion, the kinematical $2\nu$ mass-mixing parameters
appear to be strongly constrained in a basically unique region
(LMA-I), with only a marginal possibility left for the LMA-II
region. The decrease of the previous LMA-II likelihood \cite{Our1}
is an important contribution of the latest SNO data \cite{Hott}.

\section{Comments on three-family mixing}

So far, we have assumed flavor oscillations  in the active $2\nu$
channel $\nu_e\to\nu_a$  ($\nu_a$ being a linear combination of
$\nu_\mu$ and $\nu_\tau$) driven by the ($\delta m^2,
\theta_{12}$) parameters. The $(\nu_\mu,\nu_\tau)$ combination
orthogonal to $\nu_a$ is probed by atmospheric
$\nu_\mu\to\nu_\tau$ oscillations, with different parameters
$(\Delta m^2,\theta_{23})$ \cite{Revi}. As far as the third mixing
angle $\theta_{13}$ is zero (and $\delta m^2/\Delta m^2\ll 1$),
the two oscillation channels are practically decoupled, and all
our previous considerations hold without changes. However, for
small but nonzero $\theta_{13}$, the $3\nu$ survival probability
deviates from the $2\nu$ case for both solar and KamLAND $\nu_e$
oscillations:
\begin{equation}
P_{ee}^{3\nu}\simeq (1-2\sin^2\theta_{13})P_{ee}^{2\nu}\ .
\end{equation}

In particular, for $a_\mathrm{MSW}=0$,  the minimum value of
$\langle P_{ee}\rangle$ in the right panel of Fig.~2 can slightly
decrease from 1/2 to $1/2-\sin^2\theta_{13}$. Until very recently,
the upper bound on $\theta_{13}$ (dominated by CHOOZ and
atmospheric data) could be quoted as $\sin^2\theta_{13}<0.05\
(3\sigma)$ \cite{Our1,Conc}, leading to
$P_{ee}^{3\nu}(a_\mathrm{MSW}=0)>0.45$. A new SK atmospheric data
analysis \cite{EPSC}, however, appears to imply the weaker bound
$\sin^2\theta_{13}<0.067\ (3\sigma)$ \cite{Addi}, leading to
$P_{ee}^{3\nu}(a_\mathrm{MSW}=0)>0.43$. In both cases, there is no
overlap with the experimental upper bound in Eq.~(\ref{Peeupper}).
Therefore, the null hypothesis $a_\mathrm{MSW}=0$ can be rejected
at the $3\sigma$ level also in the $3\nu$ mixing case, using only
SNO(+SK) data.

In the more general case of variable $a_\mathrm{MSW}$, we have not
performed the $3\nu$ generalization of the analysis in Sec.~III.
Our educated guess is that an allowance for small values of
$\theta_{13}$ should only slightly weaken---but should not
spoil---the main results discussed therein.

\section{Conclusions}

In recent years, solar and reactor neutrino data have been shown
to be consistent with (and to favor) Mikheyev-Smirnov-Wolfenstein
effects in the flavor evolution of solar neutrinos. However, in
our opinion, the null hypothesis of ``no MSW effect'' could not be
safely rejected \cite{MSW1}. The current situation appears,
however, more favorable. In this work we have pointed out that
recent SNO data \cite{Hott} strongly favor the occurrence of MSW
effects in the solar matter and, together with world solar and
reactor data, provide a many-sigma rejection of the null
hypothesis. We have also performed an analysis where the MSW
interaction energy is freely rescaled, and found waek constraints
on the scaling parameter. These constraints can be potentially
improved by higher-statistics KamLAND data, which will then allow
more precise tests of the MSW dynamics. In the standard MSW case,
we have also performed an updated analysis of two-family active
oscillations of solar and reactor neutrinos.

We conclude by observing that, although MSW effects are an
unavoidable consequence of the standard theory of electroweak
interactions, their basic confirmation in the current neutrino
phenomenology represents an important and reassuring experimental
accomplishment, which strengthen our confidence in the emerging
picture of neutrino masses and mixings.

\acknowledgments This work is supported in part by the Istituto
Nazionale di Fisica Nucleare (INFN) and by the Italian Ministry of
Education (MIUR) through the ``Astroparticle Physics'' project. We
thank D.\ Montanino for useful discussions and suggestions.



\begin{figure}
\vspace*{-0cm}\hspace*{-2.2cm}
\includegraphics[scale=0.9, bb= 30 100 500 700]{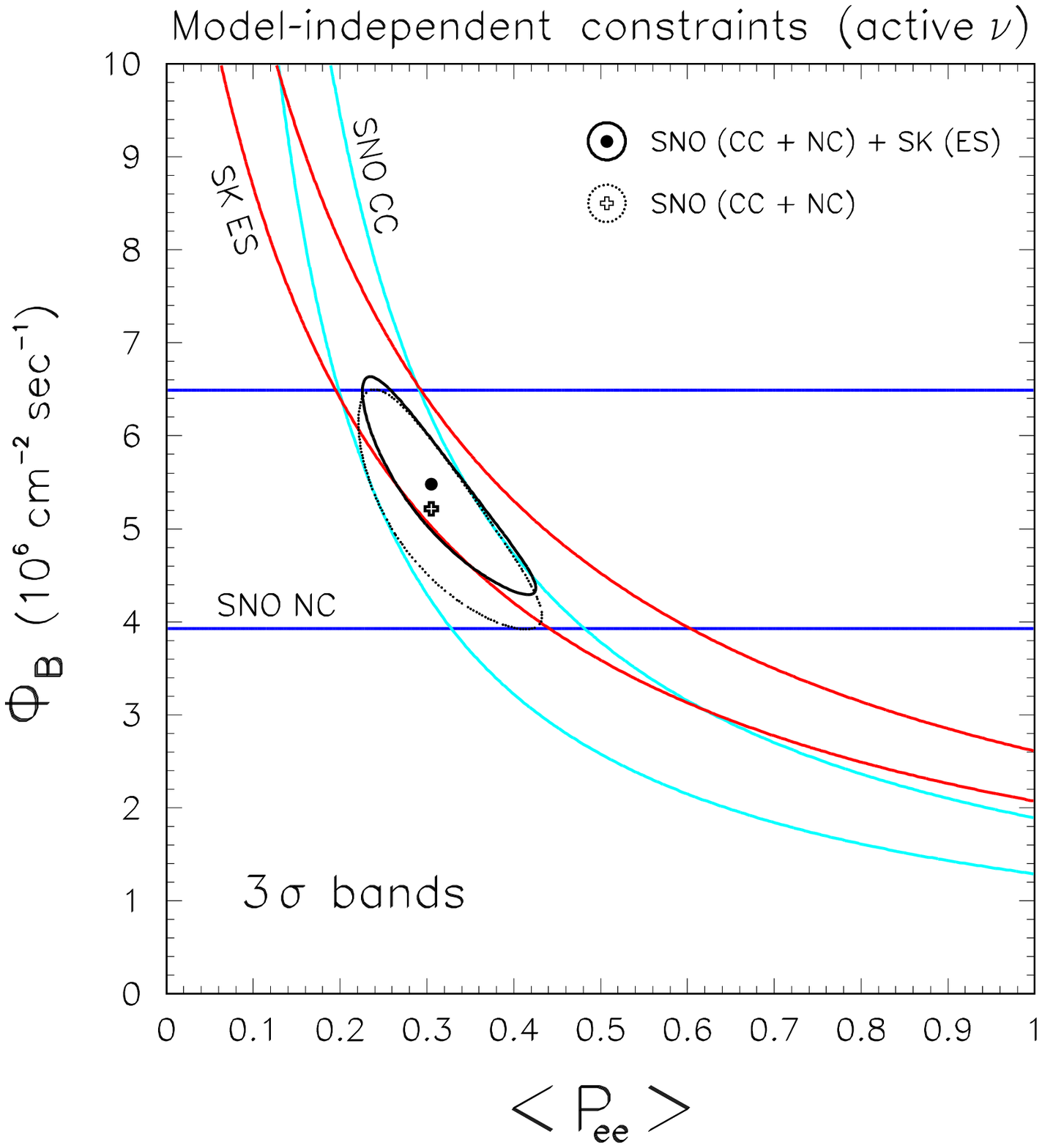}
\vspace*{-1cm} \caption{\label{fig1} Results of the
model-independent analysis of SNO (CC and NC) and SK (ES) neutrino
fluxes. The projections of the solid ellipse provide $3\sigma$
bounds on the $^8$B neutrino flux $\Phi_B$ and on the
energy-averaged $\nu_e$ survival probability $\langle
P_{ee}\rangle$.}
\end{figure}

\begin{figure}
\vspace*{-0cm}\hspace*{-2.2cm}
\includegraphics[scale=0.9, bb= 30 100 500 700]{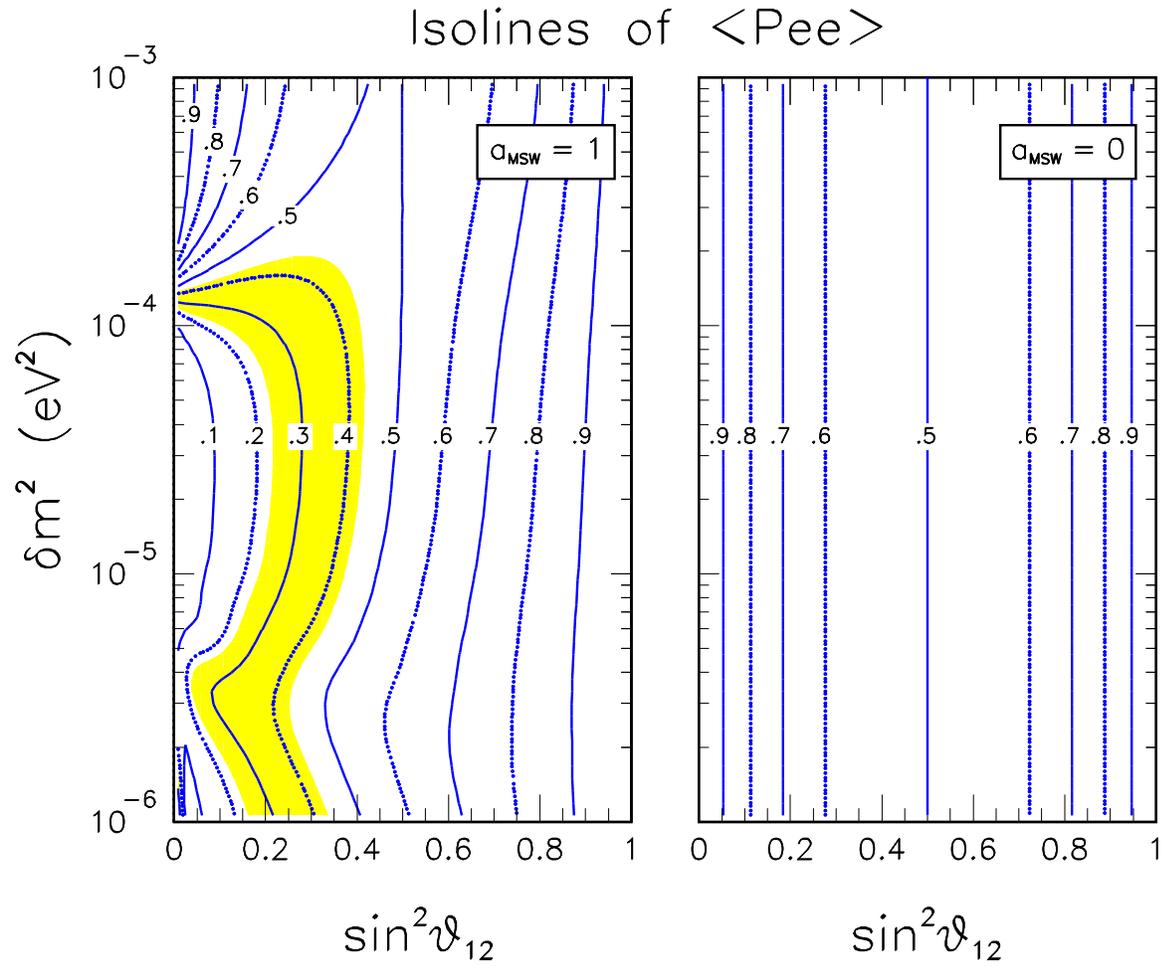}
\vspace*{-1cm} \caption{\label{fig2} Isolines of $\langle
P_{ee}\rangle$ with standard MSW effects ($a_\mathrm{MSW=1}$) and
with no matter effect ($a_\mathrm{MSW}=0$). The gray region is
allowed by the SK+SNO combination. No such region exist in the
absence of MSW effects.}
\end{figure}
\begin{figure}
\vspace*{-0cm}\hspace*{-2.2cm}
\includegraphics[scale=0.9, bb= 30 100 500 700]{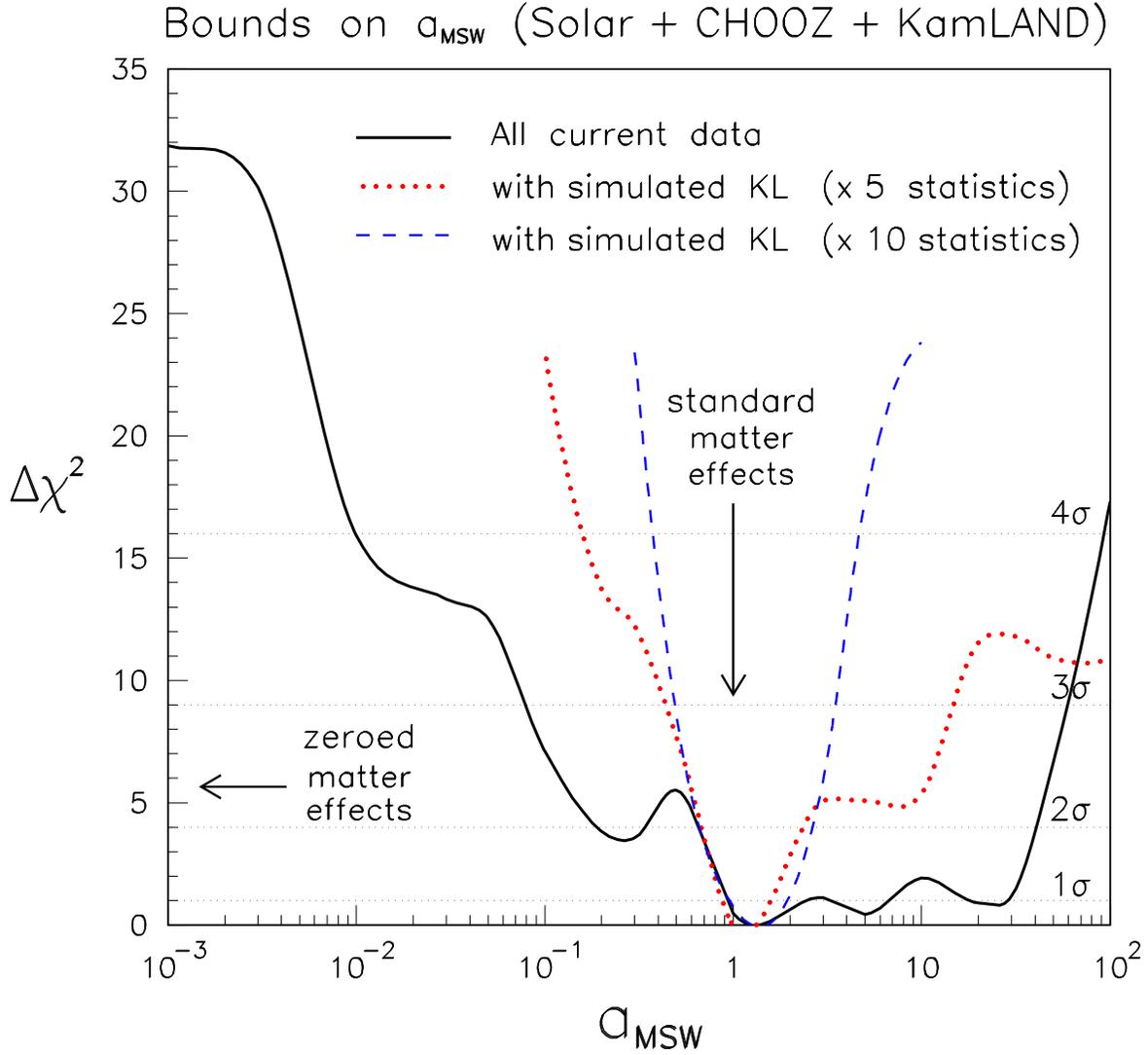}
\vspace*{-1cm} \caption{\label{fig3} Bounds on $a_\mathrm{MSW}$
(considered as a continuous free parameter), including all current
solar, CHOOZ, and KamLAND data (solid curve). Prospective KamLAND
data with higher statistics are used to draw the dotted and dashed
curves. See the text for details.}
\end{figure}
\begin{figure}
\vspace*{+2cm}\hspace*{-2.2cm}
\includegraphics[scale=0.9, bb= 30 100 500 700]{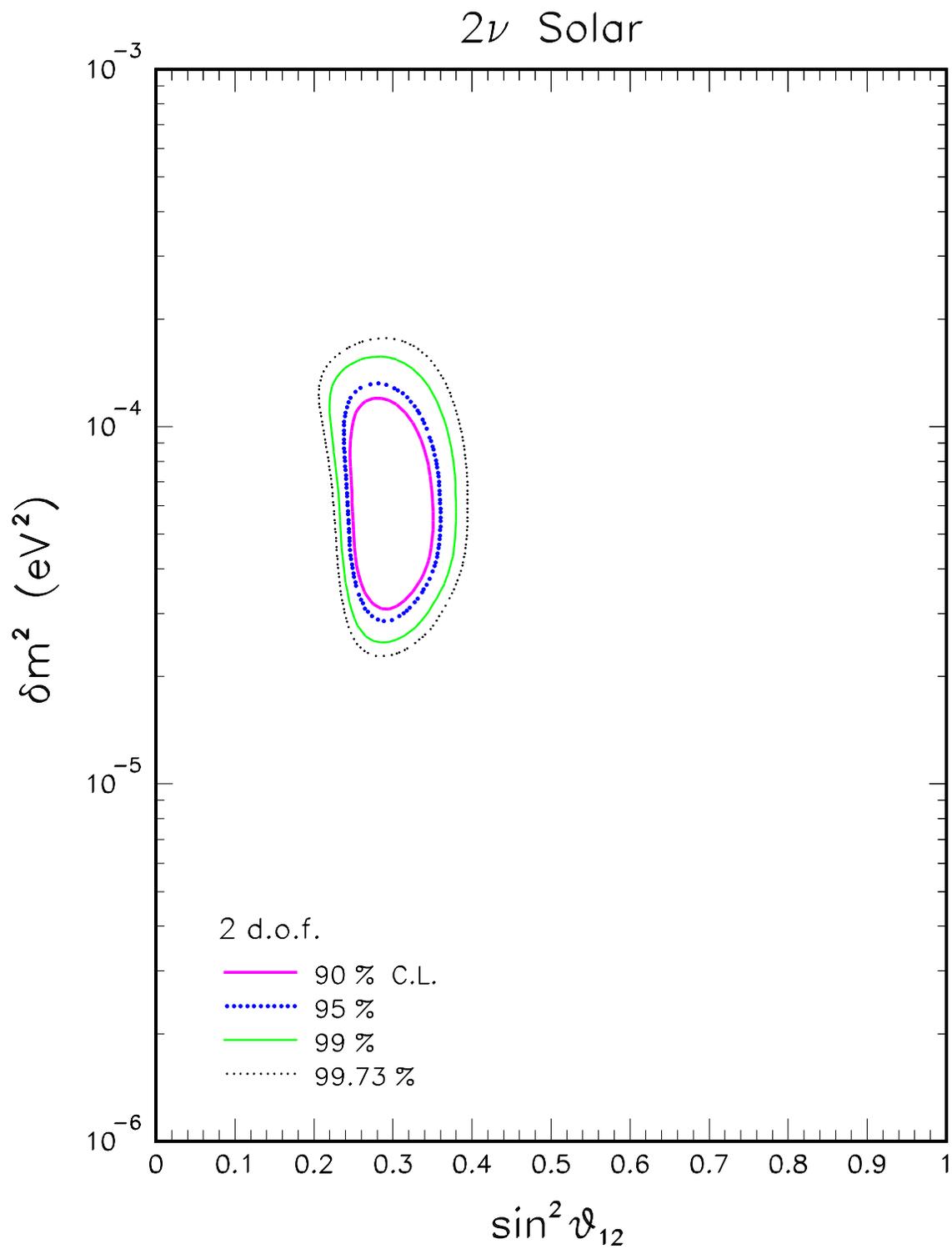}
\vspace*{0cm} \caption{\label{fig4} Results of the $2\nu$
oscillation analysis of solar neutrino data, for standard MSW
effects.}
\end{figure}
\begin{figure}
\vspace*{+2cm}\hspace*{-2.2cm}
\includegraphics[scale=0.9, bb= 30 100 500 700]{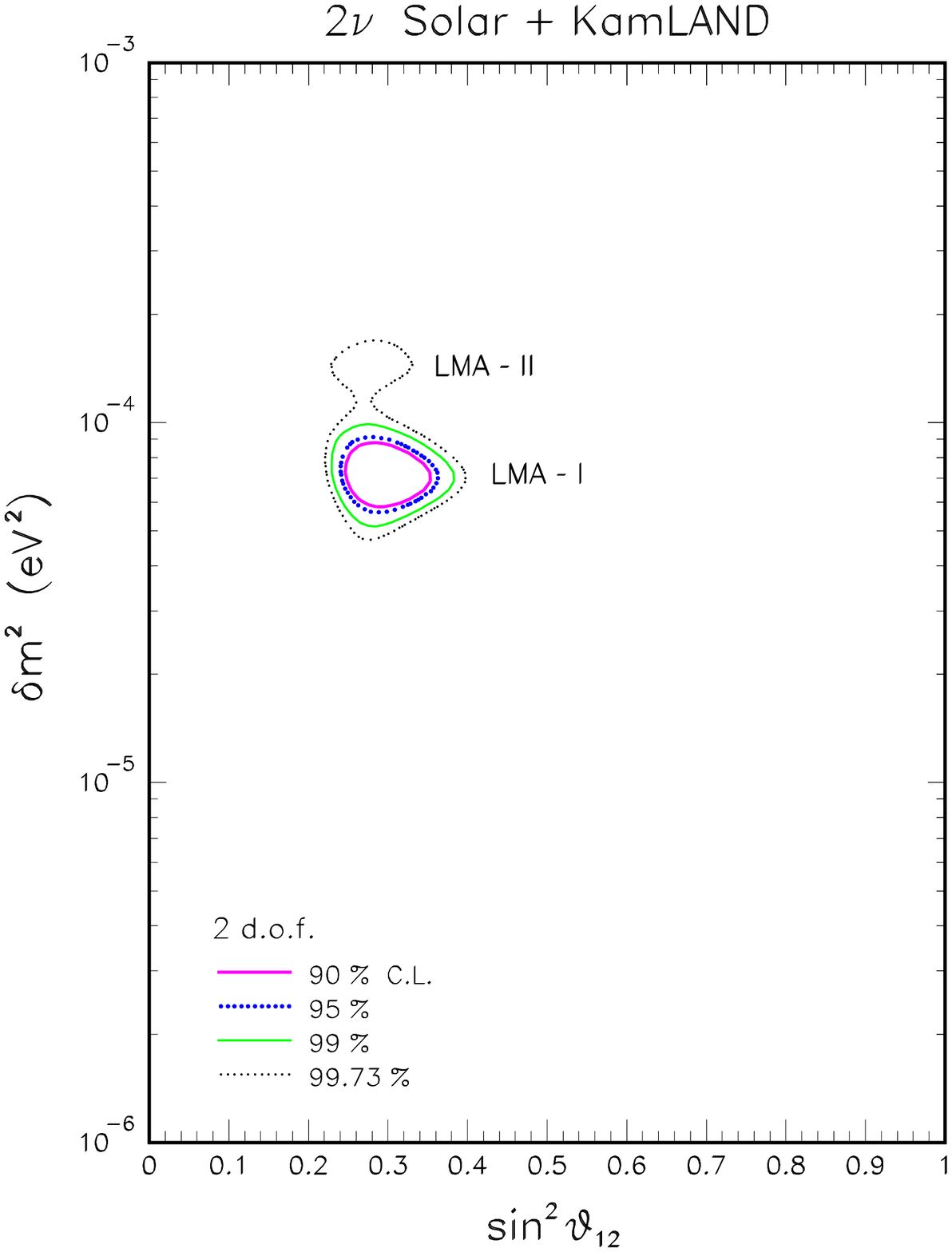}
\vspace*{0cm} \caption{\label{fig5} Results of the $2\nu$
oscillation analysis of solar and KamLAND data, for standard MSW
effects.}
\end{figure}


\begin{thebibliography}{99}

\bibitem{Hott}  SNO Collaboration, S.N.\ Ahmed {\em et al.},
                nucl-ex/0309004, submitted to Phys.\ Rev.\ Lett.

\bibitem{Book}  J.N.\ Bahcall, {\em Neutrino Astrophysics\/}
                (Cambridge U.\ Press, Cambridge, England, 1989).

\bibitem{Cl98}  Homestake Collaboration,
                B.T.~Cleveland, T.~Daily, R.~Davis Jr., J.R.~Distel,
                K.~Lande, C.K.~Lee, P.S.~Wildenhain, and
                J.~Ullman,
                Astrophys.\ J.\  {\bf 496}, 505 (1998).

\bibitem{Ab02}  SAGE Collaboration,
                J.N.~Abdurashitov {\it et al.},
                J.\ Exp.\ Theor.\ Phys.\  {\bf 95}, 181 (2002)
                [Zh.\ Eksp.\ Teor.\ Fiz.\  {\bf 95}, 211 (2002)].

\bibitem{Ha99}  GALLEX Collaboration, W.~Hampel {\it et al.},
                Phys.\ Lett.\ B {\bf 447}, 127 (1999).

\bibitem{Ki02}  T.\ Kirsten for the GALLEX/GNO Collaboration,
                in the Proceedings of {\it Neutrino 2002}, 20th International
                Conference on Neutrino Physics and Astrophysics
                (Munich, Germany, 2002), edited by F.~von
                Feilitzsch and N.~Schmitz, Nucl.\ Phys.\ B {Proc.\
                Suppl.} {\bf 118}, 33 (2003).

\bibitem{Fu01}  SK Collaboration, S.~Fukuda {\it et al.},
                Phys.\ Rev.\ Lett.\  {\bf 86}, 5651 (2001);
                {\it ibidem}, 5656 (2001).

\bibitem{Fu02}  SK Collaboration,
                S.~Fukuda {\it et al.},
                Phys.\ Lett.\ B {\bf 539}, 179 (2002).

\bibitem{AhCC}  SNO Collaboration,
                Q.R.~Ahmad {\it et al.},
                Phys.\ Rev.\ Lett.\  {\bf 87}, 071301 (2001).

\bibitem{AhNC}  SNO Collaboration,
                Q.R.~Ahmad {\it et al.},
                Phys.\ Rev.\ Lett.\  {\bf 89}, 011301 (2002).

\bibitem{AhDN}  SNO Collaboration,
                Q.R.~Ahmad {\it et al.},
                Phys.\ Rev.\ Lett.\  {\bf 89}, 011302 (2002).

\bibitem{KamL}  KamLAND Collaboration, K.~Eguchi {\it et al.},
                Phys.\ Rev.\ Lett.\ {\bf 90}, 021802 (2003).

\bibitem{CHOO}  CHOOZ Collaboration, M.~Apollonio {\it et al.}
                Phys.\ Lett.\ B {\bf 466}, 415 (1999);
                Eur.\ Phys.\ J.\ C {\bf 27}, 331 (2003).

\bibitem{Our1}  G.~L.~Fogli, E.~Lisi, A.~Marrone, D.~Montanino, A.~Palazzo and A.~M.~Rotunno,
                Phys.\ Rev.\ D {\bf 67}, 073002 (2003).


\bibitem{Pena}  J.N.~Bahcall, M.C.~Gonzalez-Garcia and C.~Pena-Garay,
                JHEP {\bf 0302}, 009 (2003).

\bibitem{Matt}  L.~Wolfenstein,
                Phys.\ Rev.\ D {\bf 17}, 2369 (1978);
                S.P.~Mikheev and A.Yu.\ Smirnov,
                Yad.\ Fiz.\ {\bf 42}, 1441 (1985)
                [Sov.\ J.\ Nucl.\ Phys.\ {\bf 42}, 913 (1985)].

\bibitem{Barg}  V.D.~Barger, K.~Whisnant, S.~Pakvasa, and R.J.N.~Phillips,
                Phys.\ Rev.\ D {\bf 22}, 2718 (1980).

\bibitem{Adia}  L.\ Wolfenstein,
                in {\it Neutrino~'78}, 8th International
                Conference on Neutrino Physics and Astrophysics
                (Purdue U., West Lafayette, Indiana, 1978), ed.\
                by E.C.\ Fowler (Purdue U.\ Press, 1978), p.~C3.

\bibitem{MSW1}  G.~L.~Fogli, E.~Lisi, A.~Palazzo and A.~M.~Rotunno,
                Phys.\ Rev.\ D {\bf 67}, 073001 (2003).

\bibitem{Vill}  F.~L.~Villante, G.~Fiorentini, and E.~Lisi,
                Phys.\ Rev.\ D {\bf 59}, 013006 (1999).

\bibitem{Gett}  G.L.~Fogli, E.~Lisi, A.~Marrone, D.~Montanino and A.~Palazzo,
                Phys.\ Rev.\ D {\bf 66}, 053010 (2002).

\bibitem{Howt}  SNO Collaboration, ``HOWTO use the SNO Salt flux
                results'', available at www.sno.phy.queensu.ca

\bibitem{BP00}  J.N.~Bahcall, M.H.~Pinsonneault, and S.~Basu,
                Astrophys.\ J.\  {\bf 555}, 990 (2001).

\bibitem{PeMa}  M.~Maris and S.T.~Petcov,
                Phys.\ Lett.\ B {\bf 534}, 17 (2002).

\bibitem{Pedr}  P.~C.~de Holanda and A.~Yu.~Smirnov,
                JCAP {\bf 0302}, 001 (2003).

\bibitem{Bala}  A.B.\ Balantekin and H.\ Yuksel, hep-ph/0309079.

\bibitem{Lown}  E. Bellotti and V. Gavrin, talks at LowNu 2003
                (Paris, France, 2003), available at cdfinfo.in2p3.fr/LowNu2003

\bibitem{Revi}  See, e.g., the reviews:
                S.M.~Bilenky, C.~Giunti, and W.~Grimus,
                Prog.\ Part.\ Nucl.\ Phys.\  {\bf 43}, 1 (1999);
                P.~Langacker, in {\it NOW 2000}, Proceedings of the
                Neutrino Oscillation Workshop 2000 (Conca Specchiulla, Italy, 2000),
                ed.\ by G.L.\ Fogli, Nucl.\ Phys.\ Proc.\ Suppl.\  {\bf 100}, 383
                (2001);
                M.~C.~Gonzalez-Garcia and Y.~Nir,
                hep-ph/0202058, to appear in Rev.\ Mod.\ Phys.

\bibitem{Conc}  M.~C.~Gonzalez-Garcia and C.~Pena-Garay,
                hep-ph/0306001.

\bibitem{EPSC}  Y.\ Hayato,
                {\em ``Status of the Super-Kamiokande, the K2K and the
                J-PARC $\nu$ project,''\/}
                talk at {\em HEP 2003},
                International Europhysics Conference on High Energy Physics
                (Aachen, Germany, 2003). Website:
                eps2003.physik.rwth-aachen.de~.

\bibitem{Addi}  G.~L.~Fogli, E.~Lisi, A.~Marrone, D.~Montanino, A.~Palazzo and
                A.~M.~Rotunno, hep-ph/0308055.




\end{thebibliography}
\end{document}